\input harvmac
\def\p{\partial}

\lref\ortin{
  N.~Alonso-Alberca, E.~Lozano-Tellechea and T.~Ortin,
  ``The near-horizon limit of the extreme rotating d = 5 black hole as a
  homogeneous spacetime,''
  Class.\ Quant.\ Grav.\  {\bf 20}, 423 (2003)
  [arXiv:hep-th/0209069].}
\lref\herd{
  G.~W.~Gibbons and C.~A.~R.~Herdeiro,
  ``Supersymmetric rotating black holes and causality violation,''
  Class.\ Quant.\ Grav.\  {\bf 16}, 3619 (1999)
  [arXiv:hep-th/9906098].}
\lref\bh{
  J.~D.~Brown and M.~Henneaux,
  ``Central Charges in the Canonical Realization of Asymptotic Symmetries: An
  Example from Three-Dimensional Gravity,''
  Commun.\ Math.\ Phys.\  {\bf 104}, 207 (1986).
}
\lref\sv{
  A.~Strominger and C.~Vafa,
  ``Microscopic Origin of the Bekenstein-Hawking Entropy,''
  Phys.\ Lett.\  B {\bf 379}, 99 (1996)
  [arXiv:hep-th/9601029].
}
\lref\dav{
  F.~David,
  ``Conformal Field Theories Coupled to 2D Gravity in the Conformal Gauge,''
  Mod.\ Phys.\ Lett.\  A {\bf 3}, 1651 (1988).
}
\lref\DistlerJT{
  J.~Distler and H.~Kawai,
  ``Conformal Field Theory And 2d Quantum Gravity Or Who's Afraid Of Joseph
  Liouville?,''
  Nucl.\ Phys.\  B {\bf 321}, 509 (1989).
}
\lref\zam{
  V.~G.~Knizhnik, A.~M.~Polyakov and A.~B.~Zamolodchikov,
  ``Fractal structure of 2d-quantum gravity,''
  Mod.\ Phys.\ Lett.\  A {\bf 3}, 819 (1988).
}
\lref\tak{
  S.~Ryu and T.~Takayanagi,
  ``Holographic derivation of entanglement entropy from AdS/CFT,''
  Phys.\ Rev.\ Lett.\  {\bf 96}, 181602 (2006)
  [arXiv:hep-th/0603001].
}
\lref\fpst{
  T.~M.~Fiola, J.~Preskill, A.~Strominger and S.~P.~Trivedi,
  ``Black hole thermodynamics and information loss in two-dimensions,''
  Phys.\ Rev.\  D {\bf 50}, 3987 (1994)
  [arXiv:hep-th/9403137].
}
\lref\asads{
  A.~Strominger,
  ``AdS(2) quantum gravity and string theory,''
  JHEP {\bf 9901}, 007 (1999)
  [arXiv:hep-th/9809027].
}
\lref\GuicaGM{
  M.~Guica and A.~Strominger,
  ``Wrapped M2/M5 duality,''
  arXiv:hep-th/0701011.
}
\lref\schmodela{
  J.~S.~Schwinger,
  ``Gauge Invariance And Mass. 2,''
  Phys.\ Rev.\  {\bf 128}, 2425 (1962).
  }
\lref\schmodelb{
    J.~B.~Kogut and L.~Susskind,
  ``How quark confinement solves the $\eta\rightarrow 3\pi$ problem''
  Phys.\ Rev.\  D {\bf 11}, 3594 (1975).
}
\lref\schmodelc{
  T.~Heinzl, S.~Krusche and E.~Werner,
  ``Nontrivial vacuum structure in light cone quantum field theory,''
  Phys.\ Lett.\  B {\bf 256}, 55 (1991)
  [Nucl.\ Phys.\  A {\bf 532}, 429C (1991)].
}
\lref\adsfirst{
  M.~Cadoni and S.~Mignemi,
  ``Entropy of 2D black holes from counting microstates,''
  Phys.\ Rev.\  D {\bf 59}, 081501 (1999)
  [arXiv:hep-th/9810251].
  M.~Cadoni and S.~Mignemi,
  ``Asymptotic symmetries of AdS(2) and conformal group in d = 1,''
  Nucl.\ Phys.\  B {\bf 557}, 165 (1999)
  [arXiv:hep-th/9902040].
}
\lref\nsn{
  J.~Navarro-Salas and P.~Navarro,
  ``AdS(2)/CFT(1) correspondence and near-extremal black hole entropy,''
  Nucl.\ Phys.\  B {\bf 579}, 250 (2000)
  [arXiv:hep-th/9910076].
}
\lref\adslast{
  M.~Cadoni and S.~Mignemi,
  ``Symmetry breaking, central charges and the \hfil\break
   AdS(2)/CFT(1) correspondence,''
  Phys.\ Lett.\  B {\bf 490}, 131 (2000) \hfil\break
  [arXiv:hep-th/0002256].
}

\Title{}{CENTRAL CHARGE FOR $AdS_2$ QUANTUM GRAVITY}

\centerline{Thomas Hartman and Andrew Strominger}
\smallskip
\centerline{Center for the Fundamental Laws of Nature}
\centerline{Jefferson Physical Laboratory, Harvard University,
Cambridge, MA 02138} \vskip .6in
\centerline{\bf Abstract}
\smallskip
Two-dimensional
Maxwell-dilaton quantum gravity on $AdS_2$ with radius $\ell$ and a constant electric field $E$ is studied.  In conformal gauge, this is equivalent to a CFT on a strip.  In order to maintain consistent boundary conditions,
the usual conformal diffeomorphisms must be accompanied by a certain $U(1)$ gauge transformation.  The resulting conformal transformations are generated by a twisted stress tensor, which has a central charge $c={3kE^2 \ell^4/4}$ where $k$ is  the level of the $U(1)$ current. This is an $AdS_2$
analog of the Brown-Henneaux formula $c = 3\ell/2G$ for the central charge of quantum gravity on $AdS_3$.
\Date{}

\listtoc\writetoc
\newsec{Introduction}
In a seminal 1986 paper Brown and Henneaux \bh\ showed that any consistent quantum theory of gravity on an asymptotically $AdS_3$ spacetime is a 2d CFT in the sense that the Hilbert space falls into a representation of the 2d conformal group.  Expressions for the Virasoro generators were presented as integrals around the boundary at spatial infinity. They further computed the Dirac brackets of these generators and found that the central charge of the CFT is
\eqn\dskl{c={3\ell_3 \over 2 G_3},}where $\ell_3$ is the $AdS_3$ radius and $G_3$ is Newton's constant.  The derivation involves only general properties of the symmetry generators and does not depend on the details of the theory. In fact an explicit unitary example of a
theory to which this result applies
was not found until a decade later \sv.

In this paper we report a similar type of result for quantum gravity with a $U(1)$ gauge field on $AdS_2$ (which is conformal to a strip). Once diffeomorphisms are fixed by imposing conformal gauge
for the metric, the residual symmetry group  is generated by one copy of the Virasoro algebra which acts non-trivially on the boundary. We argue that the central charge is
\eqn\ccg{c={3kE^2\ell^4 \over 4},}
where $\ell$ is the $AdS_2$ radius,  $E$ the electric field and $k$ is the level of the current $j_\pm$ which generates the $U(1)$.
Our arguments are fairly general and relatively insensitive to the details of the theory. We do not
have a clear example of a unitary theory to which this result applies, but we hope our result is a step in that direction.

The result \ccg\ at first seems rather surprising as it is often said (e.g.
\refs{\zam,\dav,\DistlerJT}) that  2d quantum gravity, when rewritten in conformal gauge as a CFT,
must have $c=0$ because the conformal transformations are a subgroup of the diffeomorphisms which cannot consistently have
a central term. What happens is this. For constant electric field, the  $U(1)$ potential is singular at the boundary. This has the consequence that a conformal diffeomorphism causes the potential to violate the proper boundary condition. Hence their action on the quantum Hilbert space cannot be defined.
This problem can be fixed by supplementing the conformal diffeomorphisms with a certain $U(1)$ gauge transformation generated by a current
$j_\pm$.  The resulting conformal transformation is generated by a twisted stress tensor
\eqn\trdd{\tilde T_{\pm \pm}=T_{\pm\pm}\pm{E \ell^2 \over 4}\p_\pm j_\pm.}
While the original stress tensor $T$ of necessity has $c=0$ the twisted stress tensor $\tilde T$ has central charge given by \ccg. While a $c=0$ CFT cannot be unitary, it is hence possible that some 2d quantum theory of gravity on $AdS_2$ might be recast as a unitary CFT with nonzero central charge.

This result is of potential interest for several reasons. The near-horizon geometry of every extremal black hole, including extremal Kerr,  contains a universal  $AdS_2$ factor with an electric field.\foot{For Kerr
the electric field lies in  the $SU(2)$ arising from dimensional reduction of the horizon $S^2$.} Currently, the black holes we understand microscopically are essentially those whose near-horizon region contains an $AdS_3$ factor of some kind. Hence an understanding of the near-horizon CFT with central charge \ccg\ may lead to a more universal understanding of black hole entropy, perhaps in the form of entanglement entropy along the lines discussed in \tak, \fpst.

The result is also clearly relevant to the still enigmatic $AdS_2/CFT_1$ correspondence. It has never been clear whether the dual ``$CFT_1$" should be thought of as conformal quantum mechanics or a chiral half of a 2d conformal theory. Both possibilities have been pursued. Although this is perhaps a matter of semantics, one way of distinguishing the two possibilities is that in the latter case one copy of the Virasoro generators acts nontrivially on the Hilbert space, while in the former only the global $SL(2,R)$  acts nontrivially. The approach of this paper is consistent with the idea that there is a nontrivial action of Virasoro.
In conformal gauge, 2d quantum gravity on $AdS_2$ is equivalent order by order in perturbation theory to a 2d CFT on a strip (which admits one Virasoro action). Hence $AdS_2/CFT_1$ duality becomes some kind of $CFT_2/CFT_2$ duality on the strip.

The 2d relation \ccg\ and the 3d relation \dskl\ are likely directly related by dimensional reduction. Indeed, under appropriate $S^1$ compactification
 of $AdS_3$ to $AdS_2$, the unbroken 3d conformal transformations
map to twisted 2d conformal transformations precisely as indicated by \trdd \asads.\foot{Alternately, as exploited in \refs{\herd \ortin -\GuicaGM}, the total space of the
$U(1)$ bundle describing an electric field on $AdS_2$ is $AdS_3$.}  This observation motivated the present work. However we have not succeeded in making this relation precise. It would be of great interest to do so. Part of the problem is that the appropriate $S^1$ reduction is null at the boundary of $AdS_3$, so that discrete light cone quantization must be considered.

The central charge in $AdS_2$ with a linear dilaton background was computed and related to 2d black holes in \refs{\adsfirst,\nsn,\adslast}. In that case the central charge is due to the breaking of $SL(2,R)$ by the linear dilaton.  Here we consider a constant dilaton and find a different mechanism for a non-zero central charge.

This paper is organized as follows. In section 2 we describe a particular 2d gravity effective action, which will serve as our example, and its reduction to conformal gauge. It has an $AdS_2$ solution with an electric field. In section 3 we see that conformal diffeomorphisms do not respect the boundary conditions, and show how to fix it with a $U(1)$ gauge transformation. In section 4 we describe the resulting twisted CFT and compute its central charge.

\newsec{Maxwell-dilaton gravity}

\subsec{An action}
We wish to study 2D Maxwell-dilaton theories of gravity with
an $AdS_2$ ground state supporting a constant  electric field.  While our considerations are quite general, it is pedagogically useful to have a specific example in mind. A simple example is described by the action
 \eqn\ert{\eqalign{S =
&{1\over 2\pi}\int d^2t \sqrt{-g}\left( \eta (R+{8\over \ell^2}) -
{\ell^2 \over 4}  F^2
 \right)+S_M,\cr } }
where potential boundary terms are ignored. $S_M$ is a general matter action whose effects are assumed to vanish near boundary.
It will prove convenient to introduce an auxiliary field $f$ to eliminate the quadratic gauge field term. The action then becomes
 \eqn\ermt{\eqalign{S = {1 \over 2\pi}
&\int d^2t \sqrt{-g}\left( \eta (R+{8\over \ell^2}) - {2 \over  \ell^2}  f^2+
f \epsilon^{\mu\nu}F_{\mu\nu}\right)+S_M\cr
} }
The $f$ equation of motion sets
\eqn\rts{f={\ell^2 \over 4}\epsilon^{\mu\nu}F_{\mu\nu}.}
\subsec{The vacuum solution}
This theory has an $AdS_2$ vacuum solution with
\eqn\jji{ds^2=-{\ell^2 dt^+dt^- \over (t^+-t^-)^2}}
on the Poincare wedge (where $t^\pm=t\pm\sigma$),\eqn\gfv{\bar R=-{8 \over \ell^2},} an electric
field \eqn\effl{\bar F_{+-}={2 E} \epsilon_{+-},}gauge field \eqn\rsdj{\bar
A_\pm ={E \ell^2\over 4 \sigma},}
and constant scalars
\eqn\typ{\eqalign{
\bar \eta&={E^2\ell^4 \over 4},\cr
\bar f&=-E\ell^2.}}

\subsec{CFT reformulation} We wish to study the action of the 2D conformal group on this theory by recasting it
in the form of a standard 2D CFT. To this end we
choose conformal  gauge for the metric \eqn\ddx{ds^2 =
-e^{2\rho}dt^+dt^-,}and Lorentz gauge for the $U(1)$ potential
\eqn\hoa{\p_+A_-+\p_-A_+=0.}
Locally (i.e. ignoring boundary conditions) \ddx\ fixes the coordinate system up to
residual conformal diffeomorphisms generated by
$(\zeta^+(t^+),\zeta^-(t^-))$, while \hoa\ fixes the $U(1)$ gauge up to residual transformations
generated by $\theta(t^+)+\tilde \theta(t^-)$.
In the gauge \hoa\ $A$ is determined from a scalar
\eqn\ssz{A_\pm=\pm\p_\pm a,}
so that
\eqn\dst{F_{+-}=-2\p_-\p_+a,}
and in the background solution \jji -\typ\
\eqn\frd{\bar a= {E \ell^2 \over 2 } \ln \sigma,~~~\bar \rho= -\ln \sigma +\ln {\ell \over 2 }.}

The action is then, up to total derivatives
\eqn\ertcc{\eqalign{ S =  &{1\over 2\pi}\int d^2t \bigl(
-4\p_- \eta  \p_+ \rho
+4\p_-f\p_+a
 \cr & + {4 \over
\ell^2}e^{2\rho}\eta -{1\over \ell^2 }e^{2\rho}f^2 \bigr)+S_M
. } }
The equations of motions following from \ertcc\ must as usual classically be supplemented by the gauge and gravitational constraints following from the original action \ermt:
\eqn\iik{\eqalign{G_-&=-2\p_-f+j^M_-=0,\cr  G_+&=2\p_+f+j^M_+=0,}}
\eqn\tri{\eqalign{T_{--}&=-2\p_-\eta\p_-\rho+2\p_-f \p_-a-j_-^M\p_-a+\p_-\p_-\eta+T^M_{--}=0,\cr T_{++}&=-2\p_+\eta\p_+\rho+2\p_+f \p_+a+j^M_+\p_+a+\p_+\p_+\eta+T^M_{++}=0.\cr}}
Here $T^M_{--} \equiv -{2\pi\over \sqrt{-g}}{\delta S_M \over \delta g^{--}}$
and $j^M_{-}\equiv -2\pi{\delta S_M \over \delta A_{+}}$. These constraints, together with the residual gauge freedom,  could classically be used to eliminate $\rho, ~~\eta,~~a$ and $f$ as dynamical degrees of freedom, leaving  only the matter fields as local degrees of freedom.
The equations of motion obtained by varying the fields in \ertcc\ , including the $\rho$ equation $T_{+-}=0$, imply the conservation laws\foot{In order for $T_{\pm\pm}$ to be holomorphically conserved without use of the constraints a multiple of the latter must be added to the variation of the action with respect to $g^{--}$ as in \tri. In deriving holomorphic $G$ conservation we have used
$2\pi{\p S_M\over \p a}=\p_+j^M_--\p_-j^M_+$.}
\eqn\ggdf{\p_+T_{--}=0, }
\eqn\wws{\p_-G_+=\p_+G_-=0.}
\newsec{Boundary conditions and modified conformal transformations }

In order to define the theory we must impose boundary conditions at
$\sigma =0$.
First, we wish to restrict the diffeomorphisms so that the boundary remains at $\sigma=0$. This requires
\eqn\tyi{\zeta^+(t,0)=\zeta^-(t,0).}The absence of charged current flow out of the boundary requires
\eqn\ook{\p_tf|_{\sigma=0}=0.}A well-defined variational principle then requires
\eqn\wwwm{ \p_ta|_{\sigma=0}=A_\sigma|_{\sigma=0}=0,} consistent with the background solution \frd.

A subtlety arises because the vacuum solution \frd\ diverges on the boundary $\sigma=0$: the conformal transformations \tyi\ fail to preserve the boundary condition \wwwm.  At the boundary $\p_+^n\zeta^+=\p_-^n\zeta^-$. It follows that near the boundary,
\eqn\ujh{\zeta^+(t+\sigma)-\zeta^-(t-\sigma)=2 \sigma \p_-\zeta^-(t,0)+{\cal O}(\sigma^3).}
Naively, the RHS of \ujh\ can be neglected at the boundary. However since $\bar A_\pm$ diverges we must be careful in taking the  $\sigma\to 0$ limit.
One finds the Lie derivative acts as
\eqn\ews{[{\cal L}_\zeta (\bar A_+-\bar A_-)]_{\sigma=0}= \p_+(\zeta^+\bar A_+)+\zeta^-\p_-\bar A_+-\p_-(\zeta^-\bar A_-)-\zeta^+\p_+\bar A_-={E \ell^2\over 2}\p_+^2\zeta^+(t,0).}
In order to fix this,  a diffeomorphism must be accompanied by a gauge transformation $\theta(t^+)+\tilde{\theta}(t^-)$ with
\eqn\dfu{\theta(t^+)=-{E  \ell^2\over 4}\p_+\zeta^+,~~~~\tilde \theta(t^-)={E  \ell^2\over 4}\p_-\zeta^-.}
The conformal symmetry group of the theory is hence the conformal diffeomorphisms \tyi\
supplemented by the gauge transformations \dfu. For example acting on the gauge potential we have
\eqn\rft{\delta_\zeta A={\cal L}_\zeta A-{E \ell^2\over 4}d(\p_+\zeta^+-\p_-\zeta^-).}

\newsec{The twisted CFT}

\subsec{Improving the stress tensor}

Conformal diffeomorphisms are generated via Dirac brackets with line integrals of the current ${1 \over 2\pi}(T_{++}\zeta^+,T_{--}\zeta^-)$. For the fields explicitly written in \ertcc\ these are, at fixed $t^+$
\eqn\poi{[\p_-\rho(s^-),\p_-\eta(t^-)]={\pi}\p_-\delta(s^--t^-),~~~~[\p_-a(s^-),\p_-f(t^-)]=-{\pi }\p_-\delta(s^--t^-).}
If the equations of motion imply that the
 current $j$ which generates gauge transformations  is holomorphically conserved
 \eqn\rdt{\p_-j_+=\p_+j_-=0,}
 then gauge transformations are generated via Dirac brackets with line integrals of the current ${1 \over 2\pi}(\theta j_+, \tilde \theta j_- )$, and the modified conformal transformations \rft\ are generated by
\eqn\rest{\tilde L(\zeta^-)= {1\over 2\pi}\int dt^-\tilde T_{--}\zeta^-,}
\eqn\rst{\tilde L(\zeta^+)= {1\over 2\pi}\int dt^+\tilde T_{++}\zeta^+,}
with $\tilde T_{\pm\pm}$ the improved stress tensor
\eqn\rfop{\tilde T_{\pm\pm}=T_{\pm\pm}\pm{E \ell^2 \over 4}\p_\pm j_\pm. }

\subsec{Central charge}
Let us now compute the central charge of the modified conformal transformation.  The unmodified diffeomorphisms must be an anomaly-free gauge symmetry. Therefore the Dirac bracket
\eqn\ras{\bigl[ T_{--}(t^-), T_{--}(s^-) \bigr] = -4\pi\p_-\delta(t^--s^-)  T_{--}(s^-) +2\pi\delta(t^--s^-)\p_-T_{--}(s^-) }
has no central term. On the other hand the Dirac bracket of two holomorphically conserved currents can have one\eqn\rass{\bigl[ j_{-}(t^-), j_{-}(s^-) \bigr] = -2\pi k\p_{-}\delta(t^--s^-), }
which implies that the current itself is not gauge invariant
\eqn\jjs{\delta_\theta j_-={1 \over 2\pi}[ \int \theta j, j_-]=k\p_-\theta.} The classic example of this behavior is the Schwinger model \refs{\schmodela,\schmodelb,\schmodelc}. In the fermionic formulation, the one loop diagram leads to both a central term in the commutator, as well as an anomaly in the chiral $U(1)$ current. A holomorphically conserved current of the type needed in  \rfop\ still exists but it is not gauge invariant. The anomaly and central term become classical and appear at the level of classical Dirac brackets in the bosonized Schwinger model.

Using \jjs\ one then finds from the definition \rfop\ that
\eqn\ras{\eqalign{\bigl[ \tilde T_{--}(t^-), \tilde T_{--}(s^-) \bigr] = &-4\pi\p_-\delta(t^--s^-)  \tilde T_{--}(s^-) +2\pi \delta(t^--s^-)\p_-\tilde T_{--}(s^-)  \cr
 &+{\pi kE^2  \ell^4\over 8}\p_-^3\delta(t^--s^-) .}}
Hence the modified conformal transformations have a central charge
\eqn\dsxc{c={3kE^2 \ell^4\over 4}.}
This is our main result.
\subsec{Boundary energy conservation}
Expanding around the background \jji-\typ\ with fluctuating fields $\tilde\rho,\ \tilde a,\ \tilde f,\ \tilde \eta$, the linearized equations of motion are
\eqn\psop{\eqalign{\p_+\p_-\tilde\rho &= -{1\over 2\sigma^2}\tilde\rho\cr
\p_+\p_-\tilde\eta &= -{1\over 2\sigma^2}\left(\tilde\eta + {E \ell^2\over 2}\tilde f\right)\cr
\p_+\p_-\tilde f &= 0\cr
\p_+\p_-\tilde a &= {1\over 4\sigma^2}\left( E \ell^2\tilde\rho - \half \tilde f\right).}}
The combination $\tilde\eta + {E \ell^2\over 2}\tilde f$ behaves like a massive free field with $m^2\ell^2 = 8$.  The normalizable solution falls off near the boundary as
\eqn\asp{\tilde\eta + {E \ell^2\over 2}\tilde f \sim \sigma^2.}
This in turn implies
\eqn\wwwa{\p_t\p_\sigma[\tilde\eta + {E \ell^2\over 2}\tilde f]_{\sigma=0}=0.}

Conformal invariance requires that there be no momentum flow across the boundary at $\sigma=0$, i.e.
\eqn\rsa{\tilde T_{++}(t,0)-\tilde T_{--}(t,0)=\tilde T_{t\sigma}(t,0)=0. }
This becomes
\eqn\xxv{\tilde T_{t\sigma}=-\p_t \eta \p_\sigma \tilde \rho -\p_t \tilde \rho \p_\sigma  \eta+\p_t f \p_\sigma \tilde a +\p_t \tilde a \p_\sigma f +{1 \over \sigma } \p_t \eta +{E \ell^2\over 2\sigma}\p_t f  + \p_\sigma \p_t\eta +{E \ell^2\over 2}\p_\sigma \p_t f.}
Near the boundary it follows from \wwwa\ that
\eqn\dsa{\tilde T_{t\sigma}(t,0)=0}
as required. Note that $T_{t\sigma}(t,0)\neq 0 $.

\bigskip
\centerline{\bf Acknowledgements}
We thank M. Guica, J. Maldacena, T. Takayanagi, and X. Yin for helpful discussions. This work is supported by DOE grant DE-FG02-91ER40654.
\listrefs
\end